\documentclass[a4paper]{jpconf}
\usepackage{graphicx}
\usepackage{color}

\begin{document}
\title{Study of $\eta$-meson Decays at KLOE/KLOE-2}

\author{Xiaolin Kang}

\address{
on behalf of KLOE-2 Collaboration \\
INFN-Laboratori Nazionali di Frascati, Via E. Fermi 40, 00044 Frascati (RM), Italy}

\ead{xiaolin.kang@lnf.infn.it}

\begin{abstract}
Working at a $\phi$-factory, KLOE/KLOE-2 has collected a very large and clean sample of $\eta$ decays, 
which allow to perform a variety of precision measurements.
This report presents the recent results on the $\eta$-meson decays
from KLOE/KLOE-2 experiments,
including the search of the $C$-violating decay $\eta\rightarrow\gamma\gamma\gamma$,
the determination of the $CP$ asymmetry in $\eta\rightarrow\pi^+\pi^-e^+e^-$,
the precision measurement of the $\eta\rightarrow\pi^+\pi^-\pi^0$ Dalitz plot distribution,
the search for the $CP$-violating decay $\eta\rightarrow\pi^+\pi^-$
and the precision measurement of ${\cal O}(p^6)$ dominated decay $\eta\rightarrow\gamma\gamma\pi^0$.

\end{abstract}

\section{Introduction}
After being discovered more than 50 years ago, the $\eta$ and $\eta^\prime$
mesons are still included in the physics program of various experiments,
because of their interesting properties.
Both $\eta$ and $\eta^\prime$ are eigenstates of parity $P$ and $G$, charge
conjugation $C$, and combined $CP$ parity, ($I^G(J^{PC})=0^+(0^{-+})$).
Therefore, both decays can serve as laboratories for the testing of conservation or
breaking of these discrete symmetries.
Moreover, as $\eta$ is a pseudo-goldstone boson of the strong interaction gauge theory,
its strong and electromagnetic decays are
forbidden at tree level~\cite{PST991142002} and are
particularly suitable to measure transition form
factors, investigate decay dynamics and test different ChPT models.
In addition, it is possible to search for new phenomena, such as
light dark matter bosons,
in rare or forbidden $\eta$ decays~\cite{Bbosoneta2ggp, Ubosoneta}.

The KLOE/KLOE-2 experiments~\cite{KLOEdetector} were operated at DA$\Phi$NE, an $e^+e^-$ collider
running at a center of mass energy of $\sim1020$ MeV, the mass of the $\phi$ meson,
at the INFN Frascati Laboratory.
In March 2018 the KLOE-2 experiment completed its data taking campaign
and collected 5.5 $fb^{-1}$ of data. Together with the
KLOE experiment, a total data sample of 8 $fb^{-1}$ has been collected.
Copious $\eta$ mesons are produced via the radiative decay of the $\phi$ meson,
$\phi\rightarrow\gamma\eta$, 
with a branching ratio of 1.3\%~\cite{PDG2018}.
The $\eta$ mesons can be clearly identified by their recoil against a photon
with monochromatic energy of 363 MeV.
The KLOE/KLOE-2 data sample corresponds to
$3.1\times10^{8}$ $\eta$ mesons, which allow the accurate analysis of
specific decays and to
search for rare or forbidden ones.

\section{Search for the forbidden decay $\eta\rightarrow\gamma\gamma\gamma$}

The decay $\eta\rightarrow\gamma\gamma\gamma$ is forbidden by
charge-conjugation invariance, if the weak interaction is not taken into account.
KLOE has performed a search for this decay in the data sample collected in
the years 2001 and 2002 with an integrated luminosity of 410 $pb^{-1}$
and set the upper limit ${\cal B}(\eta\rightarrow\gamma\gamma\gamma)\le1.6\times10^{-5}$
at 90\% confidence level (CL)~\cite{KLOEeta23gamma}, which is the most stringent result obtained to data.

Four isolated photons are required to select the candidate events.
A kinematic fit imposing energy-momentum conservation and time-of-flight
equal to the velocity of light is performed for the 4$\gamma$ final states,
and the $\chi^2$ value is required to be less than 25.
The dominant background is given by the process $e^+e^-\rightarrow\gamma\omega$,
where the initial-state radiation of a hard photon is followed by
the decay $\omega\rightarrow\gamma\pi^0(\gamma\gamma)$.
In addition, $\phi\rightarrow\gamma\pi^0$, $\gamma f_0(\pi^0\pi^0)$ and $\gamma a_0(\eta\pi^0)$
also can mimic four-photon
events because of the loss of photons, addition of photons from machine background
or shower splitting.
Those events can be rejected by requiring the invariant mass of
any $\gamma\gamma$ pair to lay out of the $\pi^0$ mass range, (90,180)MeV.

\begin{figure}[htb]
\begin{center}
\includegraphics[width=14pc]{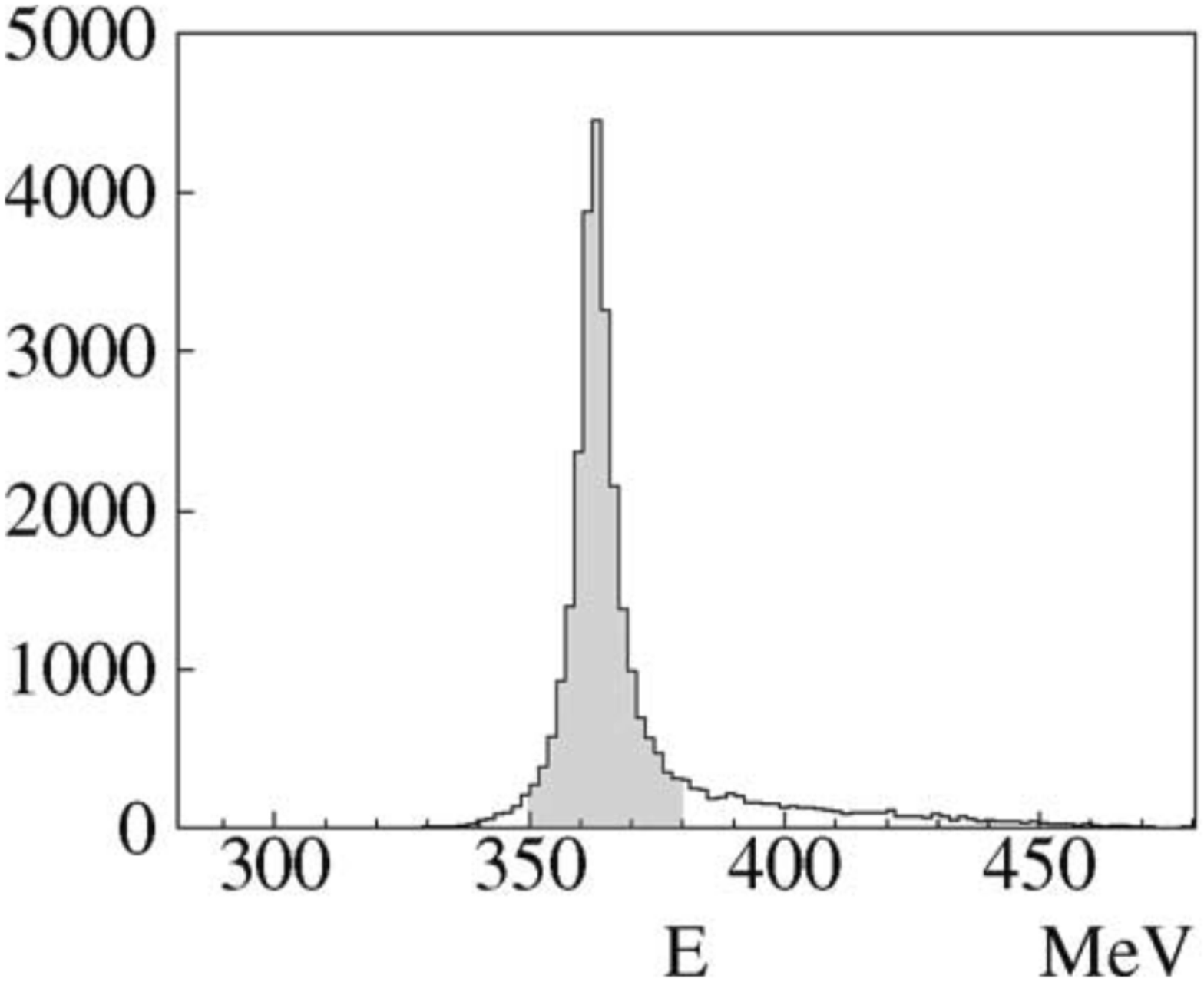}\hspace{1.5pc}
\includegraphics[width=14pc]{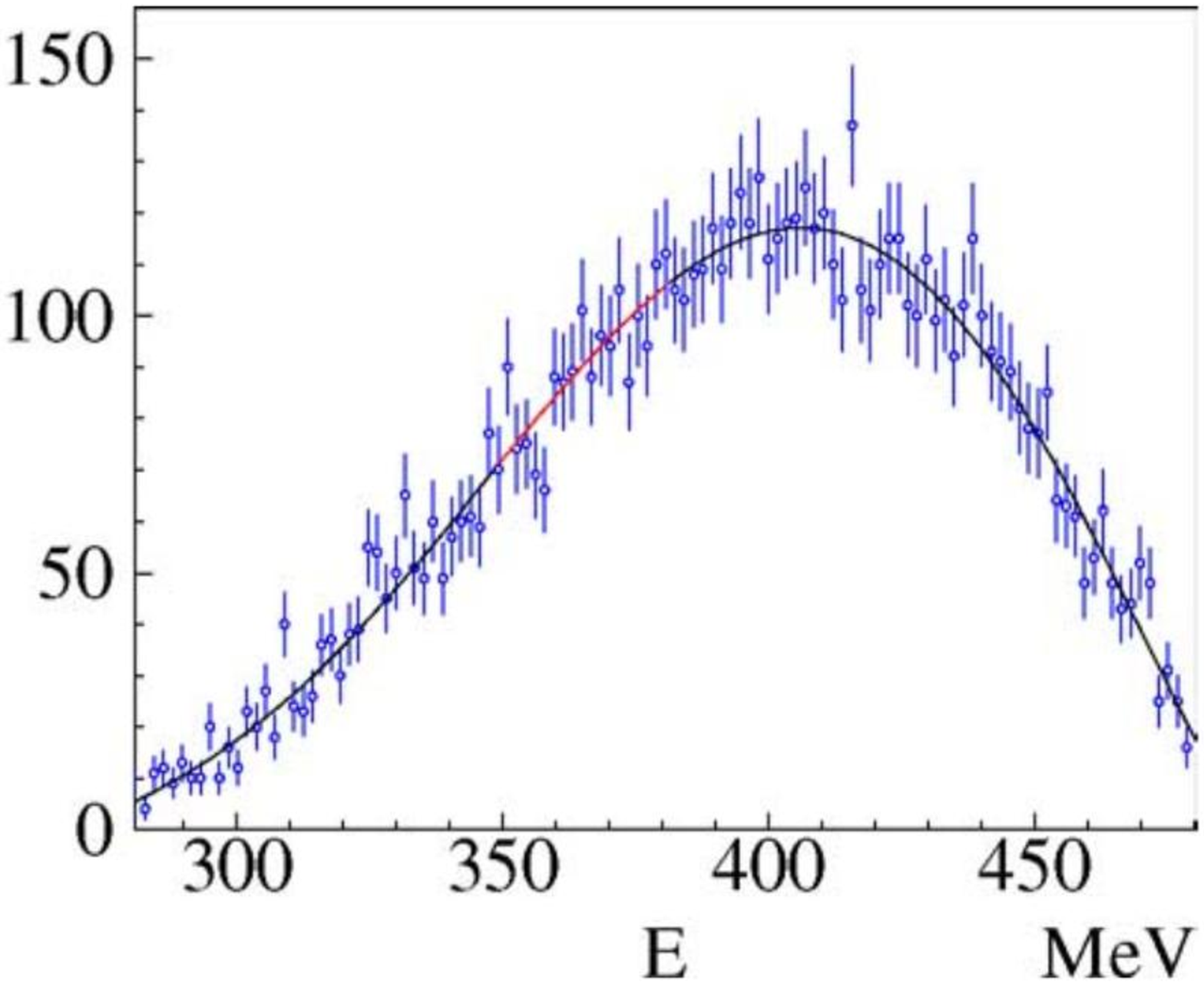}
\caption{\label{fig:3gam}Distribution of the E($\gamma_{hi}$) in the $\phi$ rest frame
for the MC simulated signal (left) and data (right). The shaded interval is
the signal region, the dots with error bars are KLOE data, the superimposed curve is the polynomial fit.}
\end{center}
\end{figure}

8268 events survive the above cuts. In the decay $\phi\rightarrow\gamma\eta$,
the recoil photon has an energy of 363 MeV in the rest frame of the $\phi$,
and it is also the most
probable energetic photon ($\gamma_{hi}$) in the signal decay chain.
Fig.~\ref{fig:3gam}-left shows the monte carlo (MC) simulated energy distribution
of the $\gamma_{hi}$, $E(\gamma_{hi})$, for the signal.
Fig.~\ref{fig:3gam}-right shows the $E(\gamma_{hi})$ distribution
for the data sample. No peak is observed in the signal region,
defined as $350<E(\gamma_{hi})<379.75$ MeV.
To estimate the background, polynomials are used to fit the
$E(\gamma_{hi})$ sidebands, (280, 350) and (379.75,481.25) MeV.
Neyman’s construction procedure is used to evaluate the upper limit,
which gives the best upper limit to date, $1.6\times10^{-5}$ @ 90\% CL.
By analysing the complete KLOE/KLOE-2 data sample, the upper limit
is expected to reach $3.6\times10^{-6}$ @ 90\% CL.

\section{$\eta\rightarrow\pi^+\pi^-e^+e^-$} 

By comparing the precision measurement of the branching ratio of
$\eta\rightarrow\pi^+\pi^-e^+e^-$ with the predictions from different
theoretical approaches, such as Vector Meson Dominance model and ChPT,
it is possible to probe
the $\eta$ meson's electromagnetic structure~\cite{PhysRep1281985}.
In addition, a possible $CP$-violating mechanism, not directly
related to the most widely studied flavor changing neutral process,
has been proposed for this decay~\cite{CPVeta2ppee}.
This mechanism could induce interference between the
parity-conserving magnetic amplitudes and the parity-violating
electric amplitudes. Such $CP$-violation effect could be tested by
measuring the polarization of the virtual photon and would result
in an asymmetry in the angle $\phi$ between pions and electrons
decay planes in the $\eta$ rest frame, defined as
\begin{center}
$A_\phi=\frac{\int_0^{\pi/2}\frac{d\Gamma}{d\phi}d\phi - \int_{\pi/2}^{\pi}\frac{d\Gamma}{d\phi}d\phi}
{\int_0^{\pi/2}\frac{d\Gamma}{d\phi}d\phi + \int_{\pi/2}^{\pi}\frac{d\Gamma}{d\phi}d\phi} =
\frac{N_{sin\phi cos\phi>0} - N_{sin\phi cos\phi<0}}{N_{sin\phi cos\phi>0} + N_{sin\phi cos\phi<0}}
$.
\end{center}
In the $\eta$ decay this asymmetry is constrained by experimental~\cite{KLOEeta2pp}
and SM~\cite{SMeta2pp} upper limits on the $CP$-violating decay
$\eta\rightarrow\pi^+\pi^-$ at the level of ${\cal O}(10^{-4})$ and
${\cal O}(10^{-15})$, respectively.

\begin{figure}[htb]
\begin{center}
\includegraphics[width=14pc]{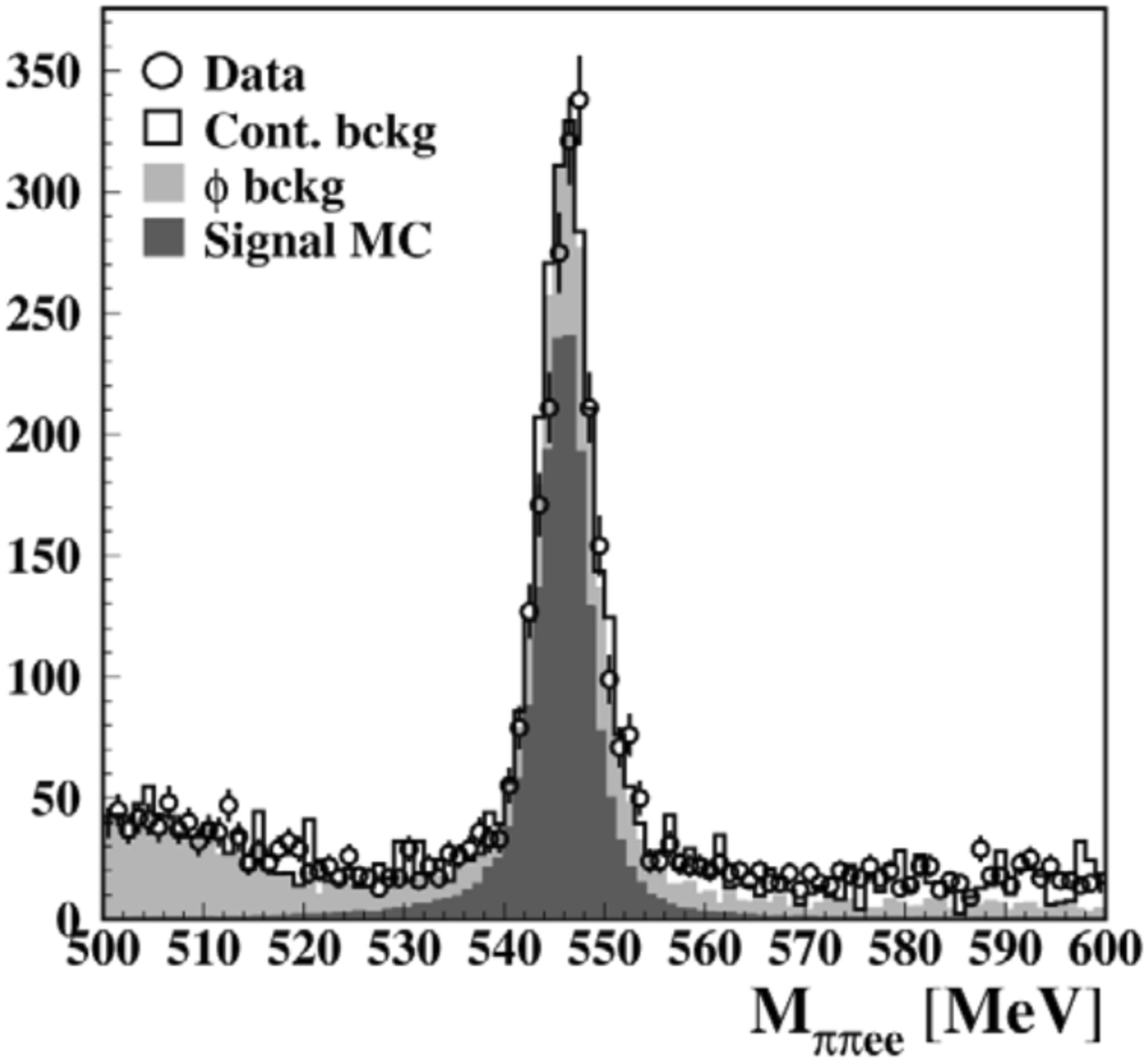}\hspace{1.5pc}
\includegraphics[width=14pc]{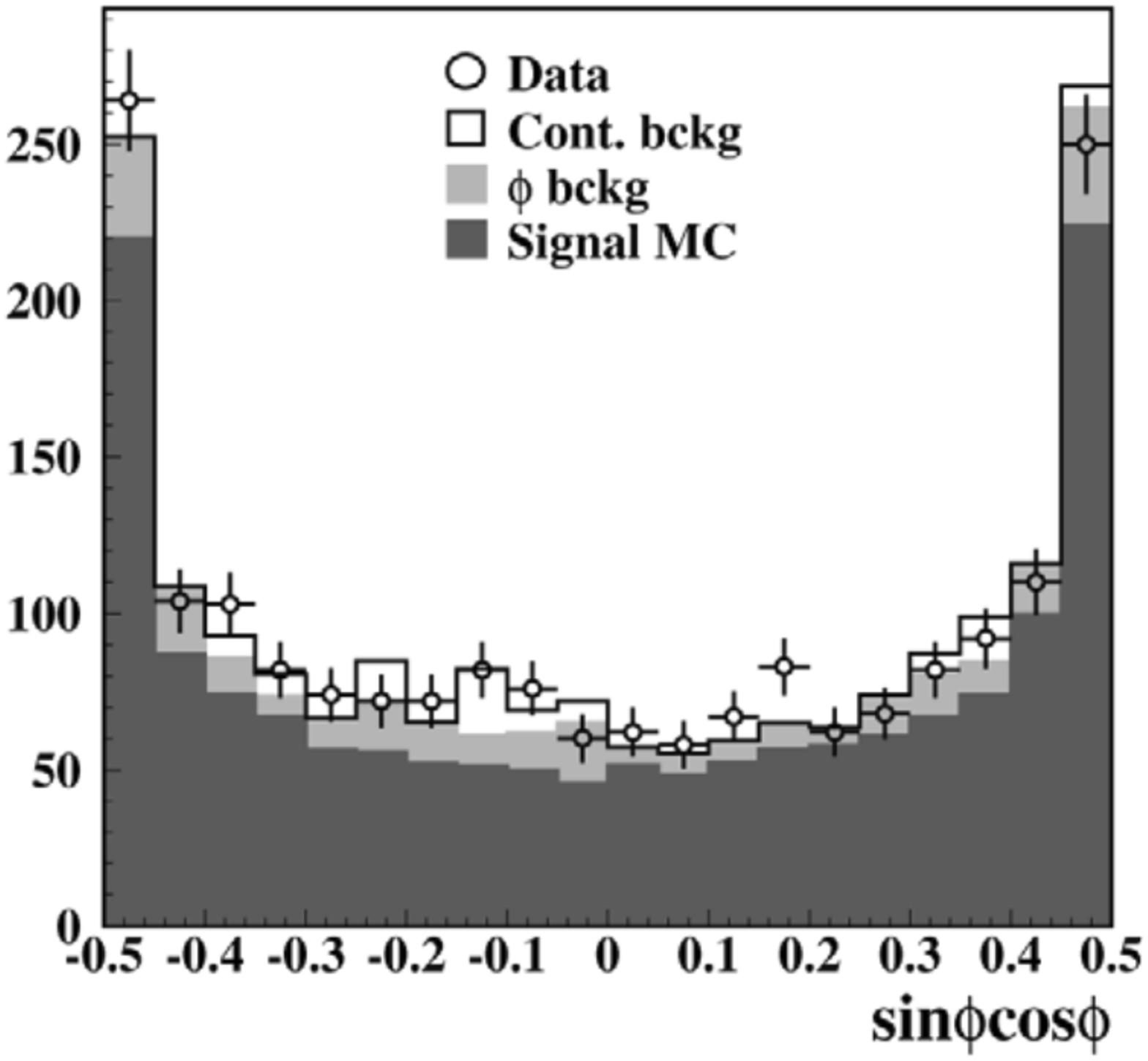}
\caption{\label{fig:eepp} $\pi^+\pi^-e^+e^-$ invariant mass spectrum around the $\eta$ mass (left) and 
the $sin\phi cos\phi$ distribution in the signal region (right).
The dots with error bars are data, the black histograms are the expected distributions obtained
by the sum of the MC contribution to signal (dark grey), $\phi$ background (light grey)
and continuum background (white).}
\end{center}
\end{figure}

KLOE has measured the $\phi\rightarrow\pi^+\pi^-e^+e^-$ decay using a
1.73 $fb^{-1}$ data sample collected at $\phi$ meson peak~\cite{KLOEeta2ppee}.
The main background sources are $\phi\rightarrow\pi^+\pi^-\pi^0$ events
(with $\pi^0$ Dalitz decay) and $\phi\rightarrow\gamma\eta$ either
with $\eta\rightarrow\pi^+\pi^-\pi^0$ (with $\pi^0$ Dalitz decay) or
$\eta\rightarrow\pi^+\pi^-\gamma$ (with photon conversion on the beam pipe).
Continuum background $e^+e^-\rightarrow e^+e^-(\gamma)$ events with
photon conversions, split tracks or interactions with some material
in the region of DA$\Phi$NE quadrupoles inside KLOE can also contaminate
the signal and are studied using off-peak data taken at $\sqrt{s}=1$ GeV.
The $\pi^+\pi^-e^+e^-$ invariant mass distribution, $M_{\pi\pi ee}$,
is shown in Fig.~\ref{fig:eepp}-left.
The background contribution is evaluated by performing a fit to the sidebands
of the $M_{\pi\pi ee}$ invariant mass distribution with the shapes of the
two background contributions obtained from MC. For the signal
estimate, the $\eta$ mass region, (535,555) MeV, is considered, counting
the event number after background subtraction. The resulting number of
signal events, $N_{\pi^+\pi^-e^+e^-}=1555\pm52$, is used to extract the
branching ratio:
${\cal B} (\eta\rightarrow\pi^+\pi^-e^+e^-(\gamma))=
(26.8\pm0.9_{stat}\pm0.7_{syst})\times10^{-5}$.

The decay plane asymmetry $A_{\phi}$ has been evaluated for the events
in the signal region after background subtraction. The obtained value is
$A_{\phi}=(-0.6\pm2.6_{stat}\pm1.8_{syst})\times10^{-2}$.
This is the first measurement of this asymmetry. The distribution of the
$sin\phi cos\phi$ variable is show in the right panel of Fig.~\ref{fig:eepp}.

\section{Dalitz plot of $\eta\rightarrow\pi^+\pi^-\pi^0$} 

The decay $\eta\rightarrow\pi^+\pi^-\pi^0$ is an isospin-violating
process. Electromagnetic contributions to the process are expected
to be very small and the decay is induced dominantly by the strong
interaction via the $u,d$ quark mass difference~\cite{Theoryeta2ppp}.
A precise study of this decay can lead to a very accurate measurement
of $Q^2=(m_s^2-\hat{m}^2)/(m_d^2-m_u^2)$, with $\hat{m}=(m_d+m_u)/2$,
and thus put a stringent constraint on the light quark masses.
The conventional decay amplitude square for $\eta\rightarrow\pi^+\pi^-\pi^0$
is parametrized as
$|A(X,Y)|^2=1+aY+bY^2+cX+dX^2+eXY+fY^3+gX^2Y+hXY^2+lX^3+\dots$,
with $X=\sqrt{3}/Q_\eta(T_{\pi^+}-T_{\pi^-})$ and
$Y=3T_{\pi^0}/Q_\eta-1$. $T$ is the kinetic energy of the pions
in $\eta$ rest frame, $Q_\eta=m_\eta-m_{\pi^+}-m_{\pi^-}-m_{\pi^0}$
is the excess energy of the reaction, $m_{\eta/\pi}$ are the nominal
masses from the Particle Data Group~\cite{PDG2018}.

\begin{figure}[htb]
\begin{center}
\includegraphics[width=20pc]{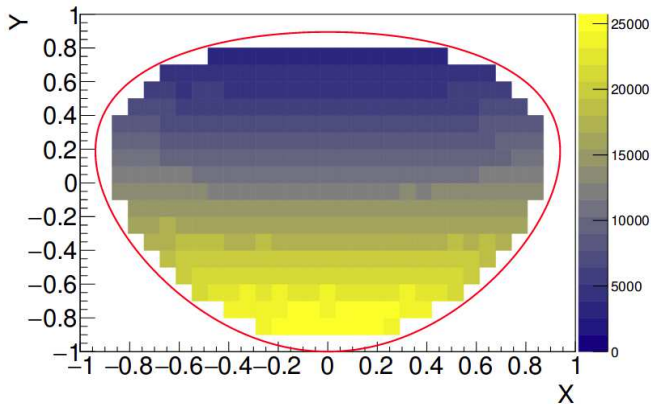}
\caption{\label{fig:eta23pi} Dalitz plot for $\eta\rightarrow\pi^+\pi^-\pi^0$
in the data sample after background subtraction.}
\end{center}
\end{figure}

The Dalitz plot distribution has been recently studied by
KLOE~\cite{KLOEeta2ppp16} with the world's largest signal
sample of $\sim4.7\times10^6$ events, based on 1.6 $fb^{-1}$
$\phi$ data, improving the previous measurement~\cite{KLOEeta2ppp08}.
From a fit to the Dalitz plot density distribution, shown in Fig.~\ref{fig:eta23pi},
we obtained the most precise determinations of the parameters that
characterize the decay amplitude. The results, together with a comparison with
recent experimental measurements, are summarized in Table~\ref{tab:eta23pi}.
This large statistics sample is also sensitive to the $g$ parameter.
As expected from $C$-parity conservation, the odd powers of $X$
($c,e,h$ and $l$) are consistent with zero.

\begin{table}[!htbp]
\caption{\label{tab:eta23pi}Recent experimental measurement of the Dalitz plot parameters
for $\eta\rightarrow\pi^+\pi^-\pi^0$.}
\centering
\begin{tabular}{lccccc}
\br
 & a & b & d & f & g \\
\mr
KLOE(16)~\cite{KLOEeta2ppp16} & $-1.095\pm0.004$ & $0.145\pm0.006$ & $0.081\pm0.007$ & $0.141\pm0.011$ & $-0.044\pm0.016$\\
KLOE(16)~\cite{KLOEeta2ppp16} & $-1.104\pm0.004$ & $0.142\pm0.006$ & $0.073\pm0.005$ & $0.154\pm0.008$ & - \\
KLOE(08)~\cite{KLOEeta2ppp08} & $-1.090\pm0.020$ & $0.124\pm0.012$ & $0.057\pm0.017$ & $0.14\pm0.02 $  & - \\
WASA~\cite{WASAeta2ppp}     & $-1.144\pm0.018$   & $0.219\pm0.051$ & $0.088\pm0.023$ & $0.115\pm0.037$ & - \\
BESIII~\cite{BESIIIeta2ppp}   & $-1.128\pm0.017$ & $0.153\pm0.017$ & $0.085\pm0.018$ & $0.173\pm0.035$ & - \\
\br
\end{tabular}
\end{table}

In addition, we have also checked the $C$-parity conservation by measuring
the left-right, quadrants and sextants charge asymmetries~\cite{CVeta2ppp},
which give the values
$A_{LR}=(-5.0\pm4.5^{+5.0}_{-11})\times10^{-4}$,
$A_Q=(+1.8\pm4.5^{+4.8}_{-2.3})\times10^{-4}$ and
$A_S=(-0.4\pm4.5^{+3.1}_{-3.5})\times10^{-4}$, respectively.
All of them are consistent with zero at the level of $10^{-4}$.

\section{Search for $\eta\rightarrow\pi^+\pi^-$} 

The decay $\eta\rightarrow\pi^+\pi^-$ violates both $P$ and $CP$ invariance.
In the SM, this decay can proceed only via the $CP$-violating weak interaction,
through mediation by a virtual $K_S^0$ meson, with a branching ratio
${\cal B}(\eta\rightarrow\pi^+\pi^-)\le2\times10^{-27}$~\cite{SMeta2pp}.
By introducing the possible
contribution of the $CP$-violating QCD $\theta$ term 
to this decay, the upper limit may reach
the order of $10^{-17}$~\cite{SMeta2pp}.
While allowing a $CP$ violation in the extended Higgs sector gives a slightly larger upper limit,
a order of $10^{-15}$~\cite{SMeta2pp}.
Any detection of larger branching fractions would indicate
a new source of $CP$ violation in the strong interaction,
beyond any considerable extension of the SM.
With the first 350 $pb^{-1}$ of data, KLOE searched for evidence of
this decay, setting the best upper limit to date,
${\cal B}(\eta\rightarrow\pi^+\pi^-)\le1.3\times10^{-5}$ at 90\% CL~\cite{KLOEeta2pp}.
Using a data sample of 1.6 $fb^{-1}$ collected at the $\phi$ meson peak,
KLOE-2 is updating this result.

\begin{figure}[!htbp]
\begin{center}
\includegraphics[width=14pc]{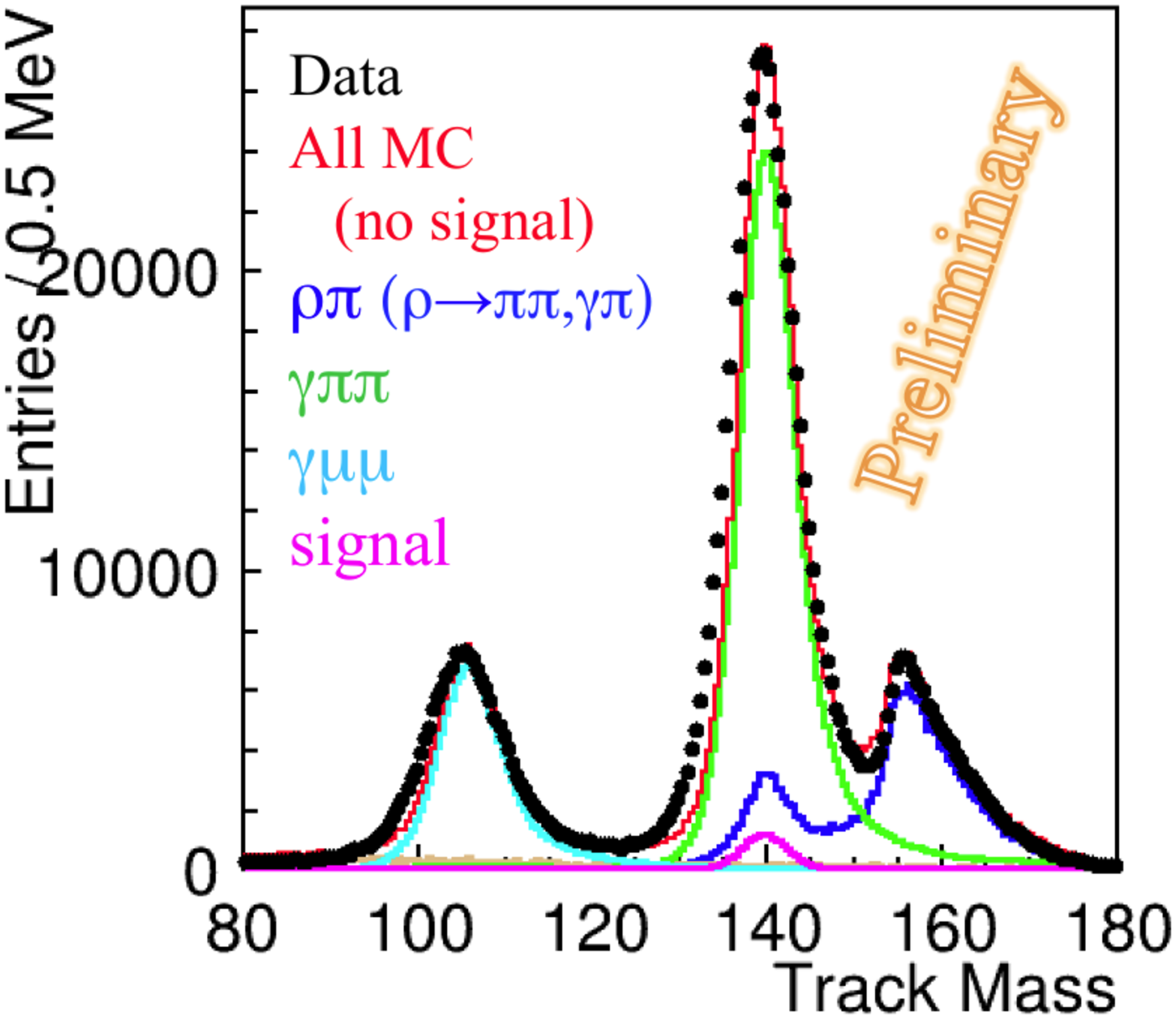}\hspace{1.5pc}
\includegraphics[width=14pc,height=12.3pc]{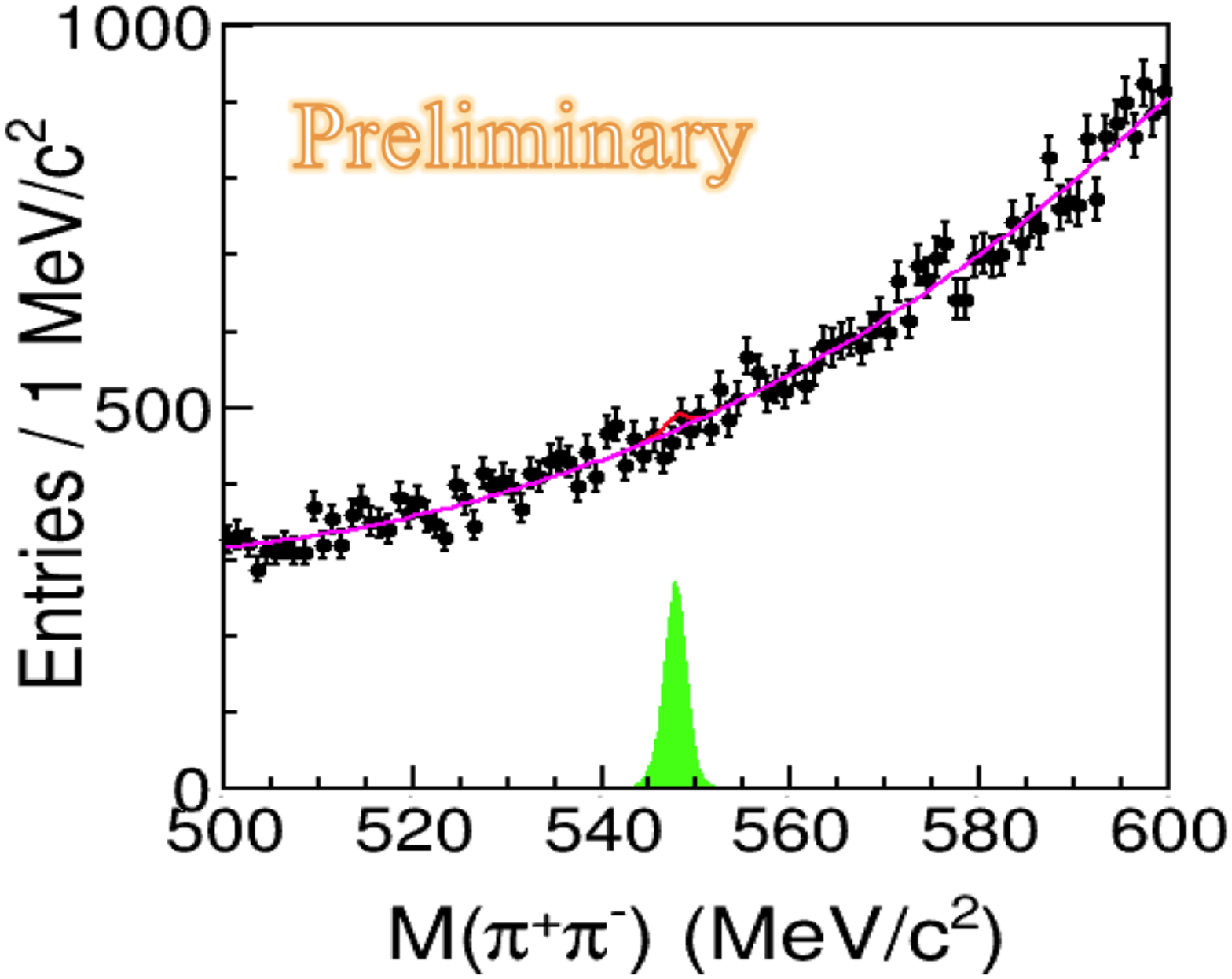}
\caption{\label{fig:eta2pp}The track-mass distribution for different decay channels (left)
and the $\pi^+\pi^-$ invariant mass spectrum between 500 and 600 MeV (right).
The signal shape in arbitrary units.}
\end{center}
\end{figure}

In this analysis, $\gamma\pi^+\pi^-$ final states are selected,
contaminated by radiative Bhabha events $\gamma e^+e^-$,
$\gamma\mu^+\mu^-$ and $\rho(\pi\pi)\pi$ with one lost photon. 
A direction match between the missing momentum of $\pi^+\pi^-$ and the
selected $\gamma$ is performed to reduce $\pi^+\pi^-\pi^0$ backgrounds.
$\gamma e^+e^-$ backgrounds can be separated from the $\gamma\pi^+\pi^-$ signal
by the different flight time inside the detector under different mass
hypothesis for each charged track.
While $\gamma\mu^+\mu^-$ events can be rejected using the so-called track-mass
variable, computed assuming the $\phi$ decays
to two particles of identical mass and a photon, shown in the left plot of Fig.~\ref{fig:eta2pp}.

The survived candidates are mainly $e^+e^-\rightarrow(\gamma)\gamma\pi^+\pi^-$.
The $\pi^+\pi^-$ mass spectra, $M(\pi^+\pi^-)$, is used to look for
the presence of $\eta$ resonance, shown in the right plot of Fig.~\ref{fig:eta2pp}.
No peak is observed in the distribution of $M(\pi^+\pi^-)$ in the vicinity of $m_\eta$.
A Bayesian method is used to obtain the upper limit of the signal number.
In the fit to $M(\pi^+\pi^-)$, the background is described by a third-order
polynomial and the signal is described with the dedicated
MC simulated shape.
The preliminary upper limit on the branching ratio is determined to be $6.3\times10^{-6}$ at 90\% CL.
With the whole KLOE and KLOE-2 data, the upper limit is expected to reach $2.7\times10^{-6}$ @ 90\% CL.

\section{The $\eta\rightarrow\gamma\gamma\pi^0$ decay} 

The doubly radiative decay $\eta\rightarrow\gamma\gamma\pi^0$ offers a  
unique window on pure $p^6$ terms of the Chiral Lagrangian,
since there is no ${\cal O}(p^2)$ contribution,
and the ${\cal O}(p^4)$ contribution is highly suppressed~\cite{ChPTeta2ggp}.
The branching ratio for this decay has been measured by several
experiments in the past~\cite{PDG2018}.
However, their results show large discrepancies.
Using a data sample of 450 $pb^{-1}$ of data, KLOE obtained the preliminary result
${\cal B}(\eta\rightarrow\gamma\gamma\pi^0)=(8.4\pm2.7\pm1.4)\times10^{-5}$~\cite{KLOEeta2ggp},
which is around $2\sigma$ lower than the predictions from Chiral-loop,
L$\sigma$M and VMD~\cite{ChPTeta2ggp,ChPTeta2ggp2} and the latest result
from Crystall Ball Collaboration at AGS~\cite{CBeta2ggp}
and A2 Collaboration at MAMI~\cite{A2eta2ggp}.
The experimental challenges of measuring this decay is 
the smallness of the rate for doubly radiative decays and
the large rate of background contamination,
in particular from $\eta\rightarrow3\pi^0\rightarrow6\gamma$ decays with lost or merged photons.
A correct estimation of the background is crucial for this analysis.

Reanalysis for this decay with more statistics is ongoing at KLOE-2, by 
performing kinematic fits under different signal or background hypothesis to suppress backgrounds.
Furthermore, multi-variable analysis with cluster shape information is under
tuning to separate single photon from merged photon.
The final measurement is forthcoming.

Moreover, this decay also provides a probability for the search of a $B$ boson at the QCD scale,
via the decay chain $\eta\rightarrow\gamma B$, $B\rightarrow\gamma\pi^0$~\cite{Bbosoneta2ggp}.

\section{Summary} 

The KLOE/KLOE-2 collaboration has achieved various
precision measurements on $\eta$-meson decays, including the tests of discrete
symmetries, studies of the decay dynamics and
searches for rare decays using the KLOE data sample.
In 2018 KLOE-2 has successfully completed its data taking at the $\phi$-meson peak.
The whole KLOE/KLOE-2 data sample amounts now to about
8 $fb^{-1}$.
Ongoing analyses will produce more precise results in the next years.

\ack
We warmly thank our former KLOE colleagues for the access to the data collected during
the KLOE data taking campaign. We thank the DA$\Phi$NE team for their efforts in maintaining
low background running conditions and their collaboration during all data taking. We
want to thank our technical staff: G.F. Fortugno and F. Sborzacchi for their dedication in
ensuring efficient operation of the KLOE computing facilities; M. Anelli for his continuous
attention to the gas system and detector safety; A. Balla, M. Gatta, G. Corradi and G.
Papalino for electronics maintenance; C. Piscitelli for his help during major maintenance
periods. This work was supported in part by the Polish National Science Centre through
the Grants No. 2013/11/B/ST2/04245, 2014/14/E/ST2/00262, 2014/12/S/ST2/00459,
2016/21/N/ST2/01727, 2016/23/N/ST2/01293, 2017/26/M/ST2/00697.

\end{document}